# Negative-Index Metamaterials: Going Optical

Thomas A. Klar, Alexander V. Kildishev, Vladimir P. Drachev, and Vladimir M. Shalaev

***Abstract*—The race to engineering metamaterials comprising of a negative refractive index in the optical range has been fueled by the realization of negative index materials for GHz frequencies six years ago. Sheer miniaturization of the GHz resonant structures is one approach. Alternative designs make use of localized plasmon resonant metal nanoparticles or nanoholes in metal films. Following this approach, a negative refractive index has been realized in the optical range very recently. We review these recent results and summarize how to unambiguously retrieve the effective refractive index of thin layers from data accessible to measurements. Numerical simulations show that a composite material comprising of silver strips and a gain providing material can have a negative refractive index of -1.3 and 100% transmission, simultaneously.**

## I. Introduction

REFRACTIVE index is the most fundamental parameter to describe the interaction of electromagnetic radiation with matter. It is a complex number $n = n' + in''$ where $n'$ has generally been considered to be positive. While the condition $n' < 0$ does not violate any fundamental physical law, materials with negative index have some unusual and counter-intuitive properties. For example, light, which is refracted at an interface between a positive and a negative index material, is bent in the "wrong" way with respect to the normal, group- and phase velocities are anti-parallel, wave and Pointing vectors are anti-parallel, and the vectors $\vec{E}$, $\vec{H}$, and $\vec{k}$ form a left-handed system. Because of these properties, such materials are synonymously called "left handed" or negative-index materials. Theoretical work on negative phase velocity dates back to Lamb (in hydrodynamics) [1] or Schuster (in optics) [2] and was considered in more detail by Mandel'shtam [3] and Veselago [4]. A historical survey referring to these and other early works can be found in Ref. [5]

In general, left handed materials do not exist naturally, with some rare exceptions like bismuth that shows $n' < 0$ at a wavelength of $\lambda \approx 60\,\mu\text{m}$ [6]. However, no naturally existing negative index material has been discovered for the optical range of frequencies. Therefore, it is necessary to turn to man made, artificial materials which are composed in such a way that the averaged (effective) refractive index is less than zero: $n'_{eff} < 0$. One a material that can display such properties photonic crystals (PC) [7-11]. However in this case, the interior structure of the material is not sub-wavelength. Consequently, PCs do not show the full range of possible benefits of left handed materials. For example, super-resolution, which has been predicted by Pendry [12], is not achievable with photonic band gap materials because their periodicity is in the range of $\lambda$. A thin slab of a photonic crystal only restores small k-vector evanescent field components because the material can be considered as an effective medium only for long wavelengths, and large k-vector components are not restored [13-15]. A truly effective refractive index $n'_{eff} < 0$ can be achieved in metamaterials with structural dimensions far below the wavelength. Metamaterials for optical wavelengths must therefore be nano-crafted.

A possible - but not the only - approach to achieve a negative refractive index is to design a material where the (isotropic) permittivity $\varepsilon = \varepsilon' + i\varepsilon''$ and the (isotropic) permeability $\mu = \mu' + i\mu''$ obey the equation

$$\varepsilon'|\mu| + \mu'|\varepsilon| < 0 . \qquad (1)$$

This leads to a negative real part of the refractive index $n = \sqrt{\varepsilon\mu}$ [16]. Equation 1 is satisfied, if $\varepsilon' < 0$ and $\mu' < 0$. However, we note that this is not a necessary condition. There may be magnetically active media (i.e. $\mu \neq 1$) with a positive real part $\mu'$ for which Eq. 1 is fulfilled and which therefore show a negative $n'$.

This work was supported in part by ARO grant W911NF-04-1-0350 and by NSF-NIRT award ECS-0210445. T. A. Klar would like to thank the Alexander von Humboldt Foundation for a Feodor-Lynen Scholarship

T. A. Klar is with the School of Electrical and Computer Engineering, Purdue University, West Lafayette, IN 47907-2035, USA, on leave from the Photonics and Optoelectronics Group, Physics Department and CeNS, Ludwig-Maximilians-Universität München, 80799 München, Germany. (e-mail: thomas.klar@physik.uni-muenchen.de)

A. V. Kildishev is with the School of Electrical and Computer Engineering, Purdue University, West Lafayette, IN 47907-2035, USA (e-mail: kildisha@ purdue.edu).

V. P. Drachev is with the School of Electrical and Computer Engineering, Purdue University, West Lafayette, IN 47907-2035, USA (e-mail: vdrachev@ ecn.purdue.edu).

V. M. Shalaev is with the School of Electrical and Computer Engineering, Purdue University, West Lafayette, IN 47907-2035, USA (e-mail: shalaev@ ecn.purdue.edu).

Up to now, we have only considered isotropic media where $\varepsilon$ and $\mu$ are complex scalar numbers. It has been shown that in the case of anisotropic media, where $\varepsilon$ and $\mu$ are tensors, a negative refractive index is feasible even if the material shows no magnetic response ( $\mu = 1$ ). If, for example, a $n' < 0$ can be achieved for an uniaxial dielectric constant with $\varepsilon_x = \varepsilon_\perp < 0$ and $\varepsilon_y = \varepsilon_z = \varepsilon_\| > 0$ [6, 17]. Despite the fact that using anisotropic media is a very promising approach, we will not focus on that topic here. This is mainly because a negative index for optical frequencies has only been achieved so far following the approach of magnetically active media.

The paper is organized as follows: In section II we recall how to achieve magnetic activity for GHz frequencies using metallic split ring resonators (SRR) and how the SRRs have been successively scaled down to shift the magnetic resonance up to THz frequencies. When the optical range is approached, the finite skin depth of metals as well as localized plasmonic resonances must be considered in addition to the simple geometric scaling of metallic structures. This opens the way to new design outlines making active use of localized plasmonic effects as it will be outlined in section III. Metamaterials containing metal nanostructures as magnetically active components usually show low transmission due to reflection and absorption. In section IV we develop an impedance-matched design to suppress reflection. In section V we add a gain material to compensate for losses and finally obtain a fully transparent layer of negative index metamaterial.

## II. Downscaling Split Ring Resonators

The first recipee how to design a magnetically active material was suggested by Pendry in 1999 [18]: Two concentric split rings that face in opposite directions and that are of subwavelength dimensions were predicted to give rise to $\mu' < 0$ (Fig. 1a). One can regard this as an electronic circuit consisting of inductive and capacitive elements. The rings form the inductances and the two slits as well as the gap between the two rings can be considered as capacitors. A magnetic field which is oriented perpendicular to the plane of drawing induces an opposing magnetic field in the loop due to Lenz's law. This leads to a diamagnetic response and hence to a negative real part of the permeability. The capacitors (the two slits and the gap between the rings) are necessary to assure that the wavelength of the resonance is larger than the dimensions of the SRR.

Very soon after that theoretical prediction, Schultz and coworkers combined the split ring resonators (SRR) with a material that shows negative electric response in the 10 GHz range and consists of metallic wires in order to reduce the charge carrier density and hence shift the plasmonic response from optical frequencies down to GHz frequencies (Fig. 1b) [19]. The outcome was the first-ever metamaterial with simultaneously negative real parts of the permeability and the permittivity [20] and consequently with a negative refractive index at approximately 10 GHz (Fig. 1c) [21, 22]. From now on the race to push left handedness to higher frequencies was open. The GHz resonant SRRs had a diameter of several millimeters, but size reduction leads to a higher frequency response. The resonance frequency has been pushed up to 1THz using this scaling technique [23, 24].

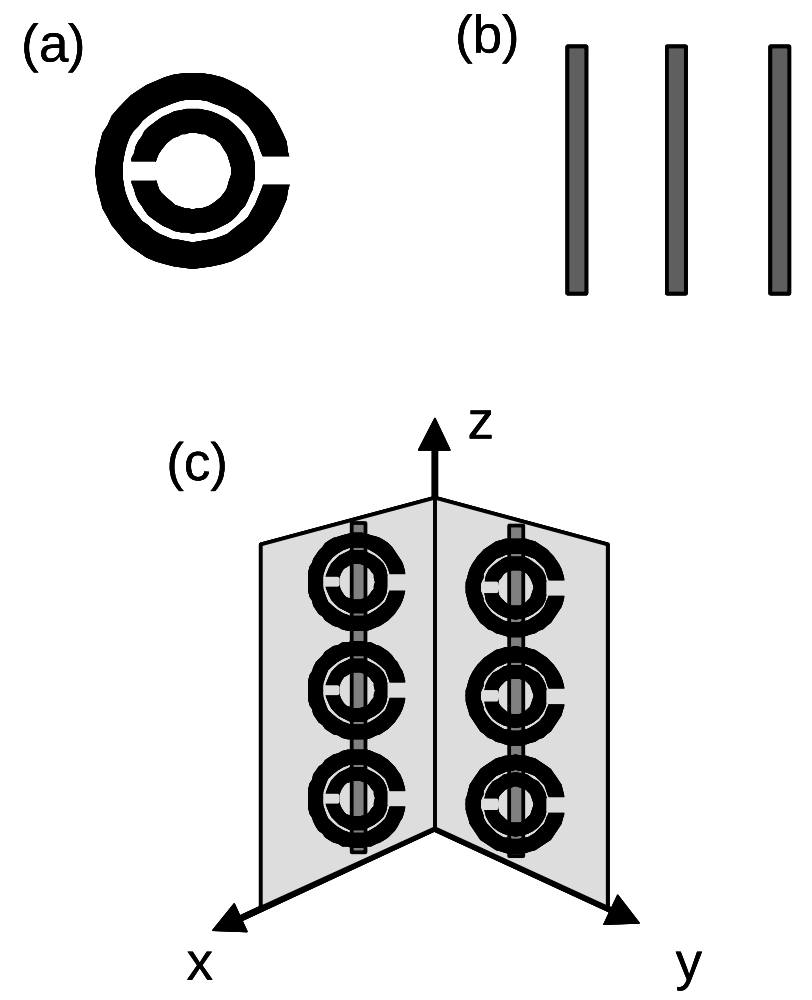

Fig. 1. (a) Magnetically resonant ( $\mu' < 0$ ) metal structure: two counter-facing split rings of sub wavelength dimensions (split ring resonator, SRR); (b) electrically resonant ( $\varepsilon' < 0$ ) metallic structure: metal rods. (c) A combination of both structures results in a negative index metamaterial $n' < 0$ .

An alternative to double SRRs is to fabricate only one SRR facing a metallic mirror and use its mirror image as the second SRR [25]. Using that technique, the resonance frequency has been shifted to 50 THz. In order to increase the frequency even more, simply downscaling of the geometrical dimensions with wavelength becomes questionable because localized plasmonic effects must be considered. However, localized plasmons open a wide field of new design opportunities. For example, a double C-shaped SRR is not required any more. Originally, the double C-shaped structure was necessary in order to shift the resonance frequency to sufficiently low frequencies such that the requirement of sub wavelength dimension could be fulfilled. In the optical range, however, localized plasmons help to shift resonance frequencies to lower energies and consequently, the doubling of the split ring is not necessary [26]. The first experimental proof that single SRRs show an electric response at 3.5 µm (85 THz) was provided in 2004 by Linden and coworkers [27] and it was concluded that the magnetic response of single SRRs should be found at the same frequency. Meanwhile the electric resonance frequencies of single SRRs has even been pushed to the important telecom wavelength of 1.5 µm [28]. Other approaches to creating metamaterials with magnetic activity that make use of localized plasmonic resonances, and abandon the classical split ring resonator shape completely, will be discussed in the following section.

## III. Metamaterials Using Localized Plasmonic Resonances

### A. Metal Nanorods

It was mentioned by Lagarkov and Sarychev [29] that a pair of noble metal nanorods can show a large paramagnetic response, and it was first pointed out by Podolskiy et al. [30] that such a pair of noble metal nanorods is also capable of a *diamagnetic response* at 1500 nm. In this publication [30], it was predicted for the first time, that materials containing such pairs of rods can show a negative $n'$ even for visible wavelengths. The issue has been discussed in more detail by Panina et al. [26] and also by Podolskiy et al. [31, 32]. It is illustrated in figure 2 how a pair of nanorods can show a negative response to an electromagnetic plane wave. Two gold rods are separated by a distance far less than the wavelength. The diameter of the cross section of the rods is also much less than the wavelength and the length of the rods may be, but does not need to be in the range of half of the wavelength. An AC electric field parallel to both rods will induce parallel currents in both rods which are in phase or out of phase with the original electric field, depending on whether the wavelength of the electric field is longer or shorter than the wavelength of the dipolar eigen-resonance of the electrodynamically coupled rods. Fig. 2a shows the induced electric dipole moment for the specific dimensions reported in [32]: a rod length of 162 nm, a diameter of 32 nm (assuming cylindrically shaped rods), and a distance of 80 nm.

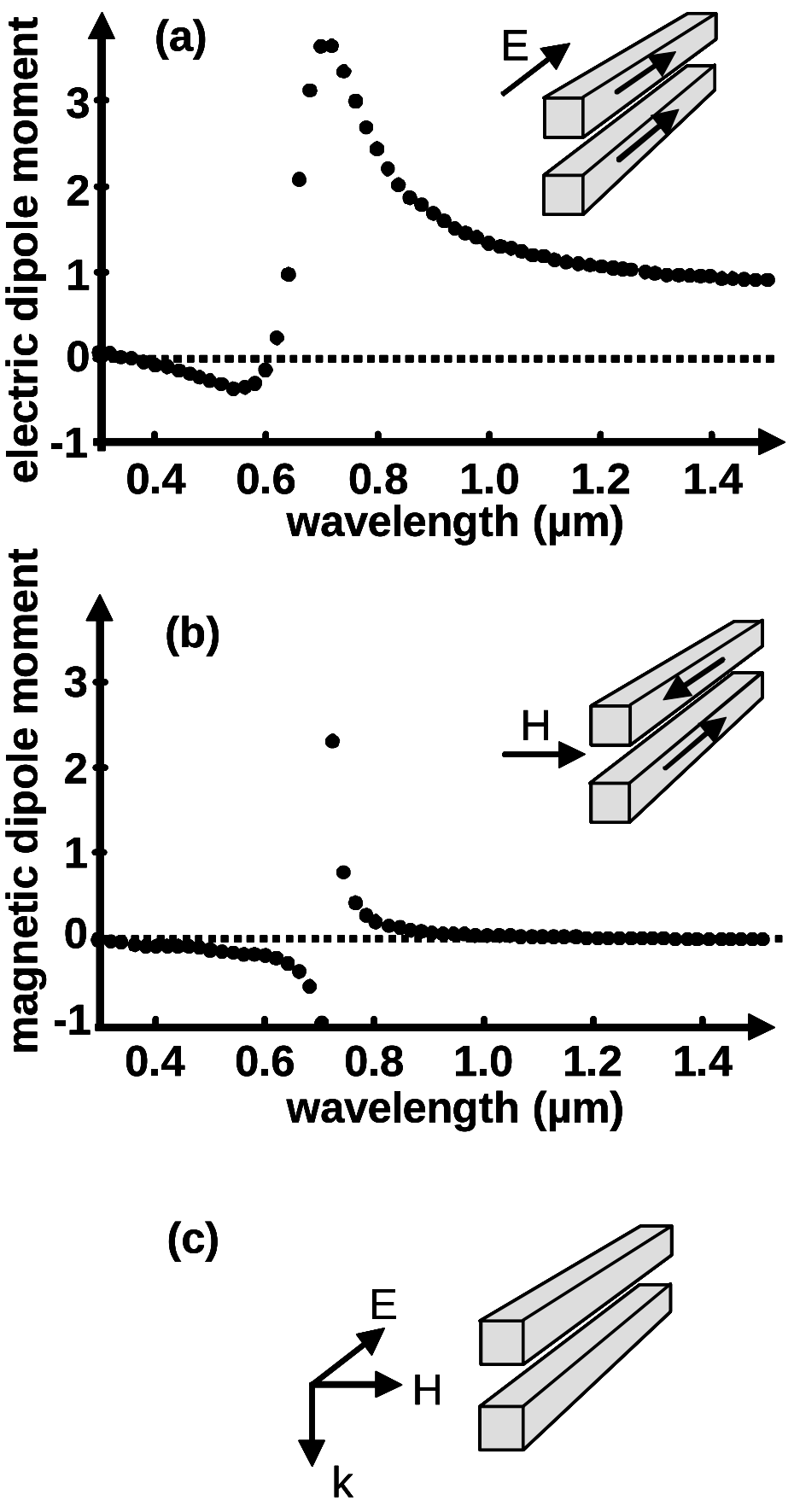

Fig. 2. Response of a pair of gold nanorods to radiation, simulated with coupled dipole approximation technique. (a) Electrical dipole moment, the electric field oriented parallel to the axis of the rods, (b) magnetic dipole moment, magnetic field oriented perpendicular to the plane of the rods, (c) a pair of rods illuminated from above with TM polarisation. The pair of rods will have a double negative response to the field.

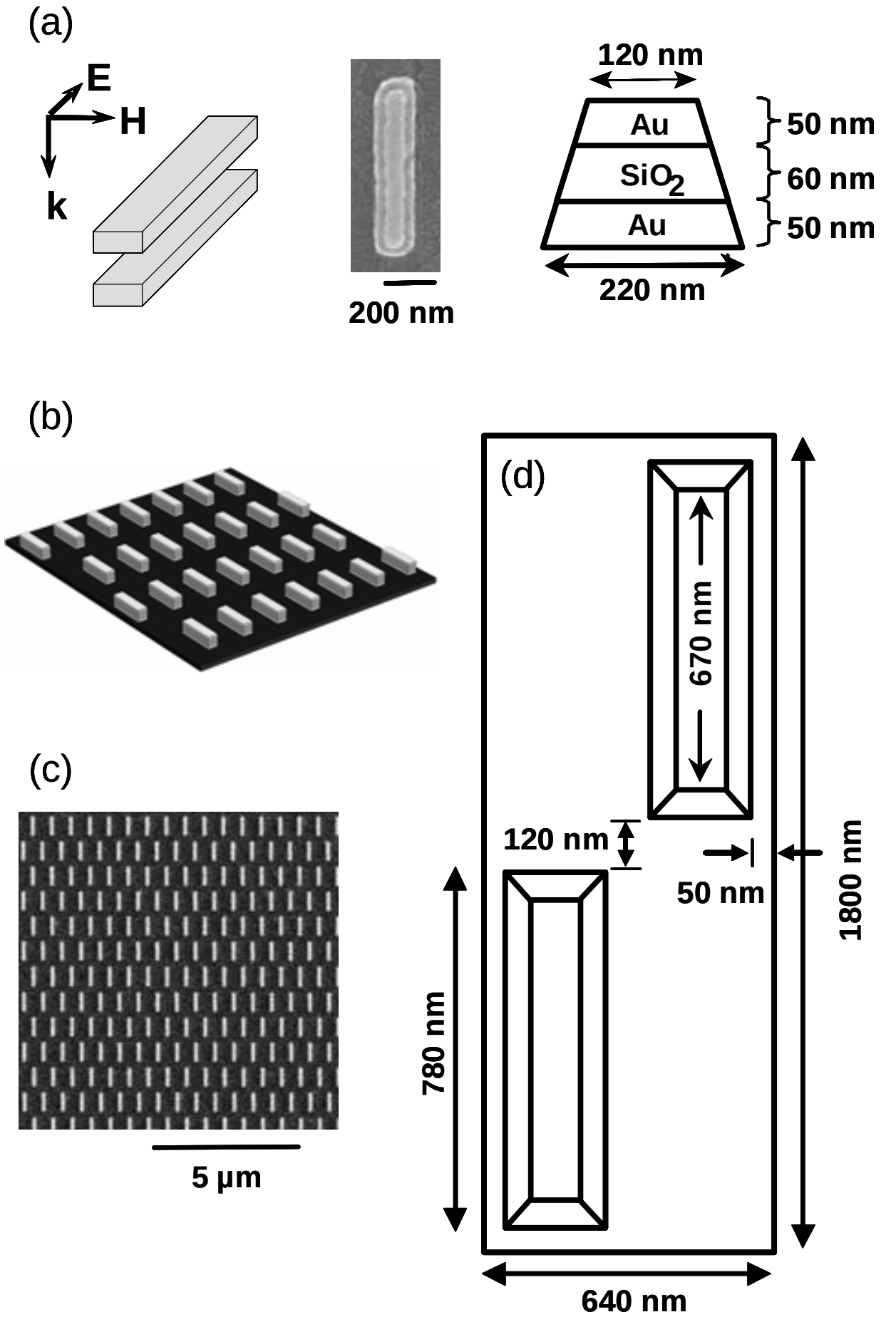

Fig. 3. (a) Left to right: scheme of nanorod pair and proper light polarization for negative index, SEM image, dimensions. (b) Scheme of the arrangement of nanorod pairs, (c) SEM image of arranged nanorod pairs. (d) Dimensions of the arrangement (one unit cell is shown).

Let us now consider the magnetic field which shall be oriented perpendicular to the plane of the rods. This magnetic field will cause anti-parallel currents in the two rods as shown in Fig. 2b. This can be considered as a dipolar magnetic mode. The magnetic response will be dia- or paramagnetic depending on whether the wavelength of the incoming magnetic field is shorter or longer than the dipolar magnetic eigenfrequency of the electrodynamically coupled rods (Fig. 2b, after [32]). In terms of coupled plasmonic resonances the magnetic dipole resonance appears at the same wavelength as the electric quadrupole resonance. However, the latter does not contribute to the electromagnetic radiation in the direction given in Fig. 2c [31].

So far, the electromagnetic response has been discussed in terms of coupled plasmonic resonances. An alternative way of looking at it is that the anti-parallel currents in the rods and the displacement currents at the ends of the two rods form a current loop or an inductance, while the gaps at the ends form two capacitors. The result is a resonant LC-circuit [29, 33].

It is important that both resonances, the dipolar electric and the dipolar magnetic resonance are at similar wavelengths. This requires that the coupling between the two rods should not be too strong, because otherwise the two resonances would be split further apart. It is seen in figures 2 a and b that there is a certain range of wavelengths (between 500 and 600 nm) where both, the induced electric and the induced magnetic dipole moments are opposing the incident fields. Hence, an electromagnetic plane wave impinging from above and with E and H oriented as shown in Fig. 2c (TM polarization) will induce a double negative response.

To the best of our knowledge, the unambiguous measurement of a negative refractive index in the optical range (specifically, at the optical telecom wavelength of 1500 nm) was reported for the first time in Ref. [34, 35]. The metamaterial in which the negative refractive index was achieved is outlined in Fig. 3. Pairs of nanorods were fabricated on a glass substrate using electron beam lithography. The actual structure of the gold nanorod doublets is shown in Fig. 3a. The nanorods are 50 nm thick, stacked on top of the glass substrate, and a 50 nm thick $SiO_2$ layer is used as a spacer. The upper rod is smaller in dimensions than the lower rod. A SEM micrograph of a single pair and its dimensions is shown in Fig. 3a. Pairs of nanorods are periodically repeated as depicted in Fig. 3b and shown by a SEM micrograph in Fig. 3c. Figure 3d shows the unit cell of the periodic arrangement and gives more dimensions. A full description of the sample and its preparation, is given in [35-37]. A similar sample containing pairs of gold nanoparticles show reflection spectra that can be explained if a negative

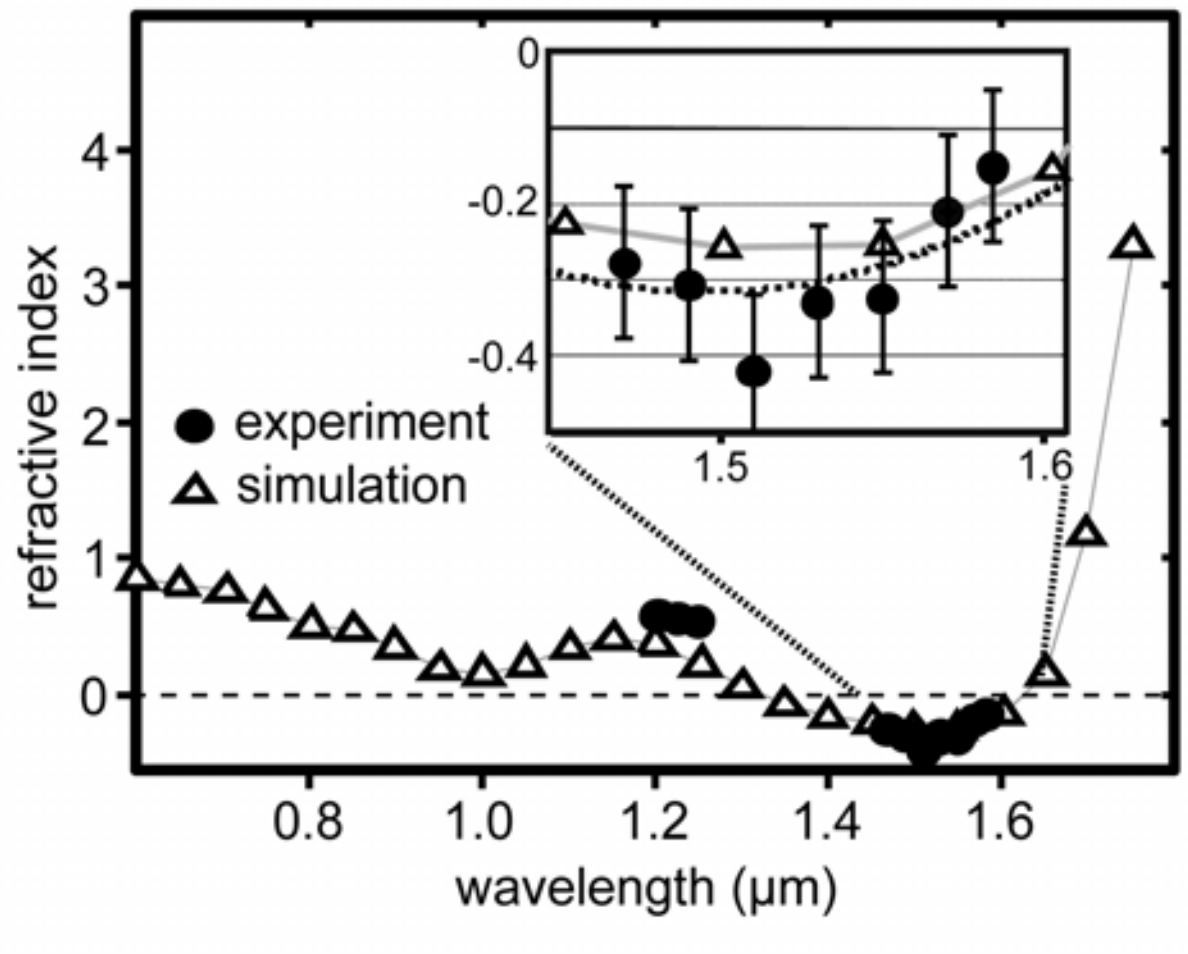

Fig. 4 The real part of the refractive index of a layer of nanorod pairs as shown in Fig. 3. Full circles represent data which are restored from experimentally determined transmission, reflection, and phase measurements. Open triangles: FDTD simulation. Inset: zoom of the region of negative refraction. The dashed line is a least square fit to the experimental data. A refractive index of $n' = -0.3 \pm 0.1$ was determined.

permeability is assumed [38]. However, a negative refractive index was not observed in that work.

Figure 4 shows the results obtained in [35] for the real part of the refractive index of the metamaterial shown in Figure 3. The full circles show experimental results and the open triangles give the results as obtained from simulations using the finite difference method in time domain (FDTD). It is clearly seen that the real part of the refractive index becomes negative in the wavelength range from approximately 1400 nm to 1600 nm, which includes the important telecommunication band at 1500 nm. The inset gives a closer look to that frequency range. The experimental data proves that $n' = -0.3 \pm 0.1$ was obtained in [35].

It turns out to be non trivial to experimentally determine the exact value of the refractive index for a thin film. In the present case, the film of negative refraction was only 160 nm thick. Therefore, the straightforward method of determining $n$ by applying Snell's law to the incoming and refracted beams can not be used. A different method to unambiguously determine the refractive index requires the measurement of the transmission $T$, the reflectance $R$ and the absolute phases of the transmitted and reflected electric fields $\tau$ and $\rho$, respectively. If those four quantities are measured, the refractive index $n = n' + in''$ in a thin, passive $(n'' > 0)$ film sandwiched between air (top) and a glass substrate (bottom) can be determined uniquely as it has been discussed in [37, 39] using transfer matrices:

$$n = \frac{1}{k\Delta} \arccos \frac{1 - r^2 + n_s t^2}{\left[1 + n_s - (1 - n_s) r\right] t}, \tag{2}$$

where $k = 2\pi/\lambda$ is the wave vector of light, $\Delta$ is the thickness of the thin film, $n_s$ is the refractive index of the glass substrate, and $r$ and $t$ are the complex reflection and transmission coefficients:

$$t = \sqrt{T}\, e^{i\tau}, \; r = \sqrt{R}\, e^{i\rho}. \tag{3}$$

The importance to measure $\tau$ and $\rho$ in addition to $T$ and $R$ was emphasized in [36] where it was shown that two similar, but not identical, samples of pair of nanorods can show the same values $T$ and $R$ (within experimental error) but greatly differ in $\tau$. Consequently, one sample was found to have negative $n'$ while for the other one $n'$ is positive.

Figure 5a shows the transmission and reflection spectra of the negative index metamaterial of Fig. 3. In order to measure the absolute phase, the beam of a tunable semiconductor laser was split in two orthogonally polarized beams, where one beam passed through the negative index metamaterial of thickness $\Delta$ while the other beam was used as a reference and passed only through the glass substrate at a spot not covered by the metamaterial [36], (Fig. 5b). The beams were recombined behind the glass substrate. The phase difference

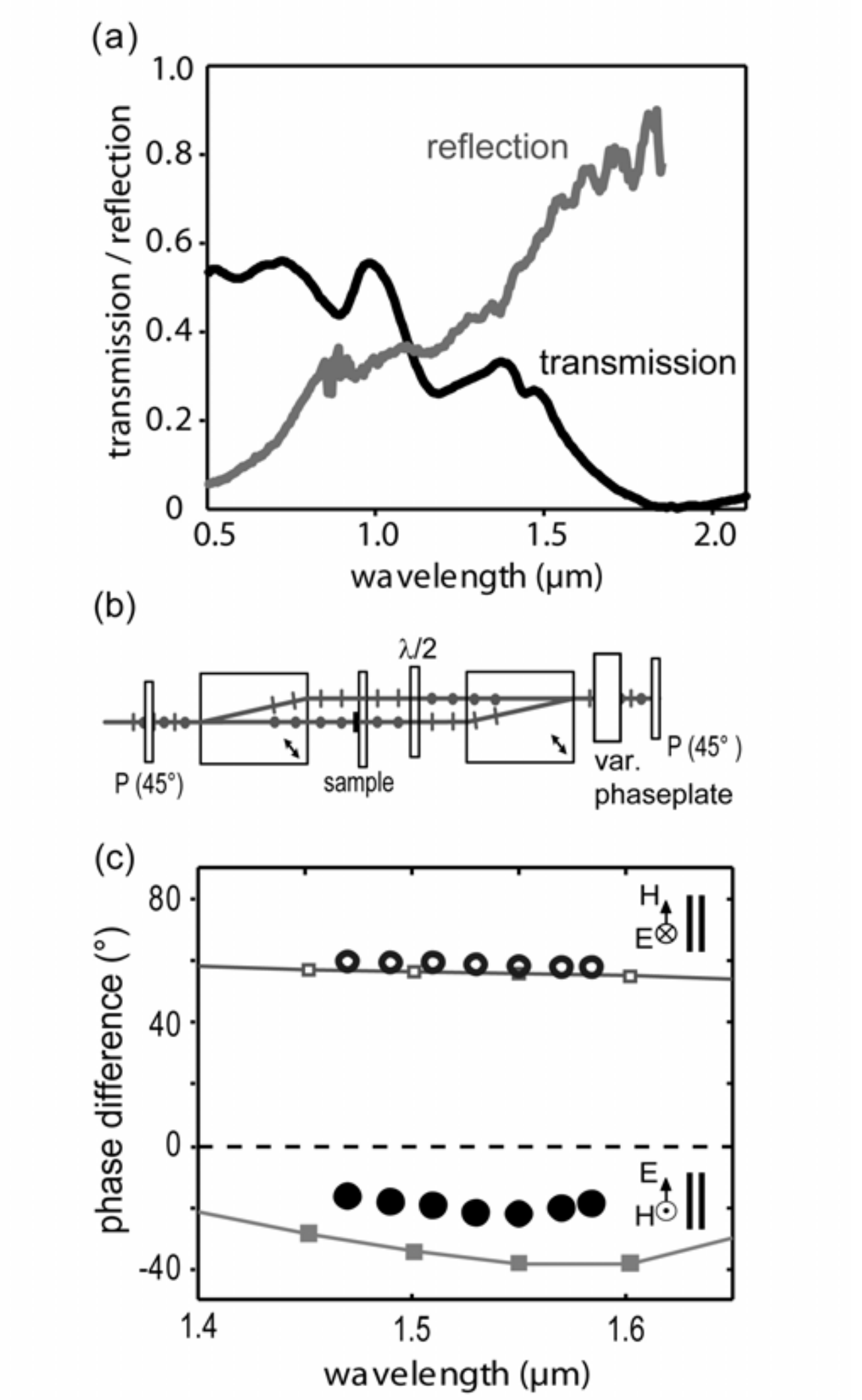

Fig. 5 (a) Measured transmission and reflection spectra of the sample shown in Fig. 3. (b) Setup for phase measurements. (c) Phase difference in the two light paths as shown in(b). Circles are measured values, squares and lines are from simulation. The light is delayed in case of TE polarisation (H-field parallel to rod pair, open symbols). In contrast, the phase is advanced in the case of TM polarisation.

$\tau$ between the beam passing through the thin film and the reference beam propagating only through air of the same thickness $\Delta$ was determined using interferometry (Fig. 5c). The phase shifts in reflection ($\rho$) were obtained for both polarizations in a similar way.

In [35] it was found that the phase $\tau$ is delayed in the metamaterial by approx. 60° compared to air in case of TE polarization (electric field perpendicular to the plane of rods). In contrast, $\tau$ is advanced by approx. 20° in case of TM polarization (Fig.5c, Ref. [35]). The advancement of $\tau$ for TM polarization can be used as an indirect evidence of $n' < 1$. However, to unambiguously prove that $n' < 0$, the complete set $(T, R, \tau, \rho)$ must be obtained, so that $n$ can be reconstructed using Eq. 3 [37].

Nevertheless, one can use pure phase measurements to make an estimate for $n'$ as it was pointed out in [37]. In the case of low reflection ($R << 1$), the following equation holds:

$$n' \approx \frac{\tau}{k\Delta}, \tag{4}$$

while in the limit of strong reflection ($R \approx 1$) the following equation holds:

$$n' \approx \frac{\tau - \rho - \frac{\pi}{2}}{k\Delta}. \tag{5}$$

These two formula indeed give an upper and lower bound to the correct value of $n'$ according to equations (2) and (3) (see Fig.6) [37].

*B. Voids*

An interesting approach to negative index metamaterials is to take the inverse of a resonant structure [40], e.g. a pair of voids as the inverse of a pair of nanorods [41-43]. The basic idea is illustrated in Fig. 7a. Instead of a pair of metal nano-ellipses separated by an oxide, which are similar to the pair of rods in Fig. 2, two thin films of metal are separated by an oxide and mounted on a glass substrate. Then, an elliptically shaped void is etched in the films (Fig. 7a, right hand side), thus forming the negative of the original paired metal ellipse structure (Fig. 7a, left hand side). Both samples should have similar resonance behavior if the orientation of the electric and magnetic fields are also interchanged. FDTD simulations were performed to determine the refractive index of void metamaterials [37]. The dimensions were chosen according to Fig. 7b in the simulations in order to match the dimensions of the experimental sample reported in [43].

These simulations were carried out for both cases of polarizations: the electric field oriented along the long axis of the elliptical voids and perpendicular to it. It is seen that $n'$ becomes negative in both cases, however, the effect is more pronounced if the electric field is oriented along the short axis (Fig. 7c). Further more, at approximately 1600 nm the real part of $n$ is negative while the imaginary part is less than 1 indicating lower losses compared to the double rod sample discussed before, where the imaginary part of the refractive index was 3 [35]. Experimental measurements with samples similar to those sketched in Fig. 7a, but with spherical voids instead of elliptical voids, confirmed a negative $n'$ at a wavelength of 2 µm [41]. The imaginary part $n''$ is large in that case, however it has been shown that further optimization can reduce $n''$ substantially [43].

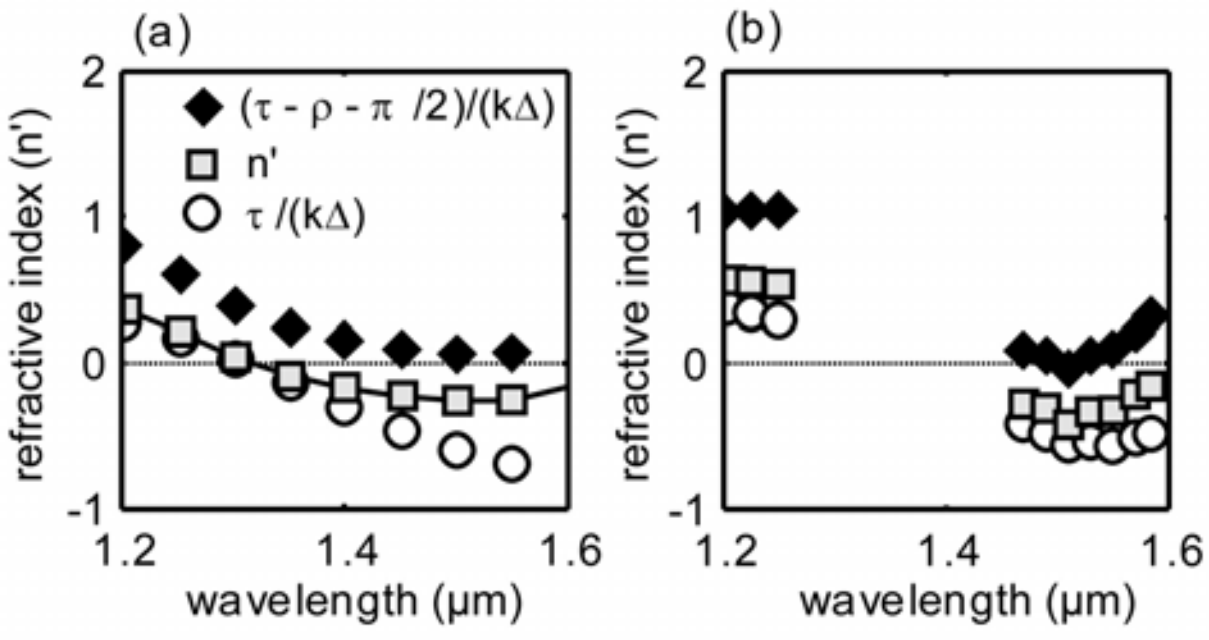

Fig. 6 Real part of the refractive index as determined by the exact formula (Eq. 2) (squares) or by phase only assumptions according to Eq. 5 (full diamonds) or Eq. 4 (open circles). (a) numerical simulations, (b) experimental results.

## IV. Pairs of Metal Strips for Impedance Matched Negative Index Metamaterials

Metamaterials using plasmon resonant metal nanoparticles have two distinct problems, each of them reducing the overall transmission through the metamaterial: The first one is absorptive losses (in terms of a large $n''$), because ohmic losses are generally large due to the excitation of localized plasmon resonances in the nanostructures. A possible solution to this problem will be discussed in the next section. In this section we will concentrate on the second issue, which is impedance matching. The impedance is given by $Z^2 = (Z' + iZ'')^2 = \mu\varepsilon^{-1}$ and it is required that the impedances match at a boundary between two media in order to eliminate reflection. This condition is well known for microwaves and replaces Brewster's law for optical frequencies if $\mu \neq 1$ [26]. Impedance is matched at a boundary between a negative index metamaterial and air, if $Z' \to 1$ and $Z'' \to 0$ in the metamaterial.

In Figure 8a we introduce a metamaterial where the conditions $Z \to 1 + 0 \cdot i$, $n' < -1$, and $n'' < 1$ hold simultaneously for a visible wavelength. The structure consists of pairs of coupled silver strips. Both strips are 280 nm wide (x-direction), 16 nm thick, and they are infinitely long in the y-direction. The two silver strips are separated in the z-direction by a 65 nm thick layer of $Al_2O_3$. The pairs of strips are periodically repeated in the x-direction with a period of 500 nm. We assume air above and below the layer of strips. In our finite element frequency domain (FEMFD) simulations this layer of metamaterial is illuminated from above with plane waves at normal incidence (along the z-direction). The electric field is polarized in the x-direction. The magnetic field, which is parallel to the strips, induces anti-parallel currents in the two silver strips as indicated in the magnified inset of Fig. 8a by the two white arrows. This leads to a magnetic response of the structure. We use FEMFD

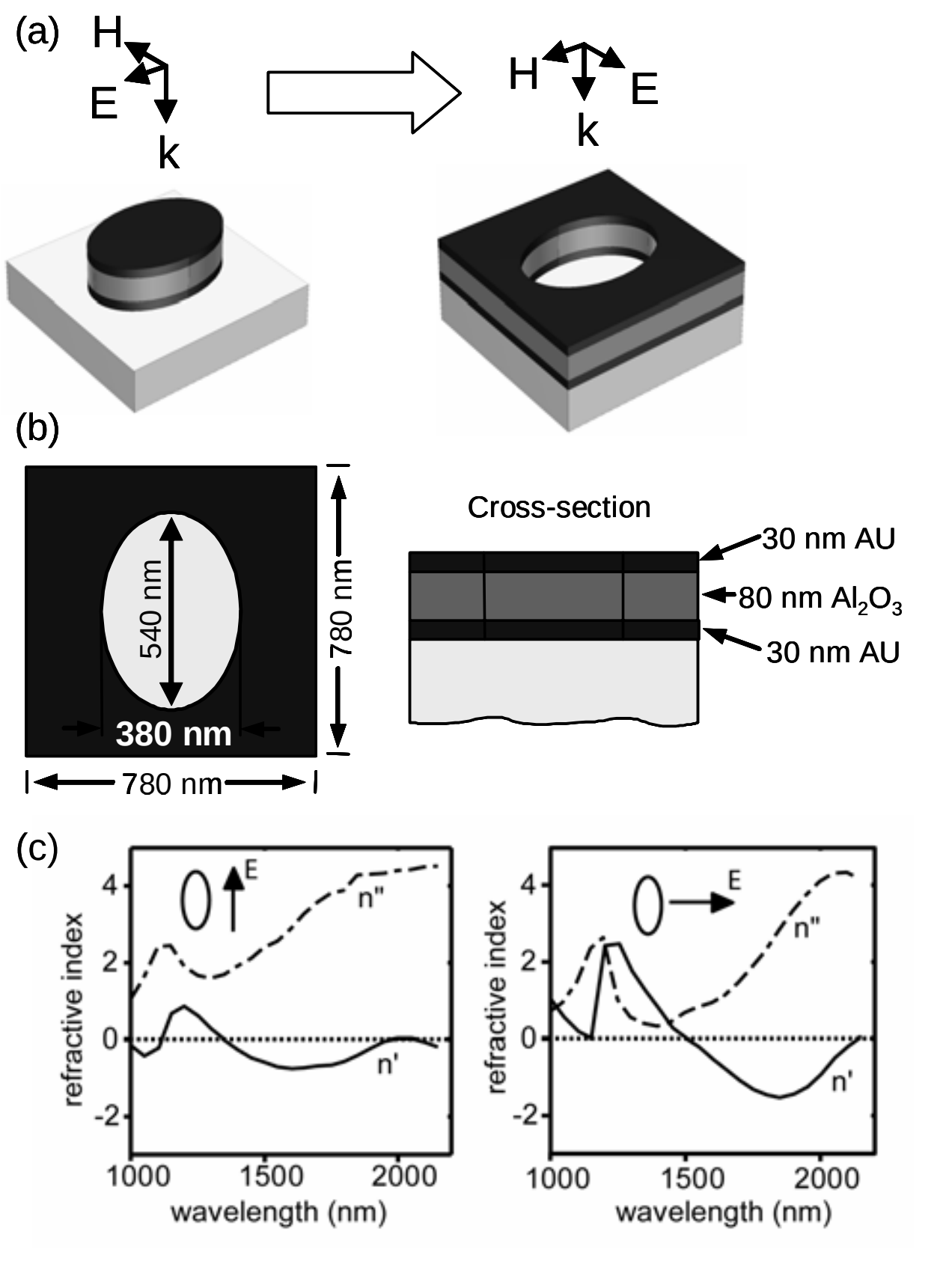

Fig. 7 (a) Left: Nano-ellipse consisting of two 30 nm thick ellipses of gold separated by 80 nm of $Al_2O_3$. Right: An elementary cell of coupled elliptic voids. (b) Dimensions of the voids. The voids are repeated periodically in 2D. (c) Refractive index $n = n' + in''$ for light polarized parallel (left) or perpendicular (right) to the long axis of the voids as obtained from FDTD simulations.

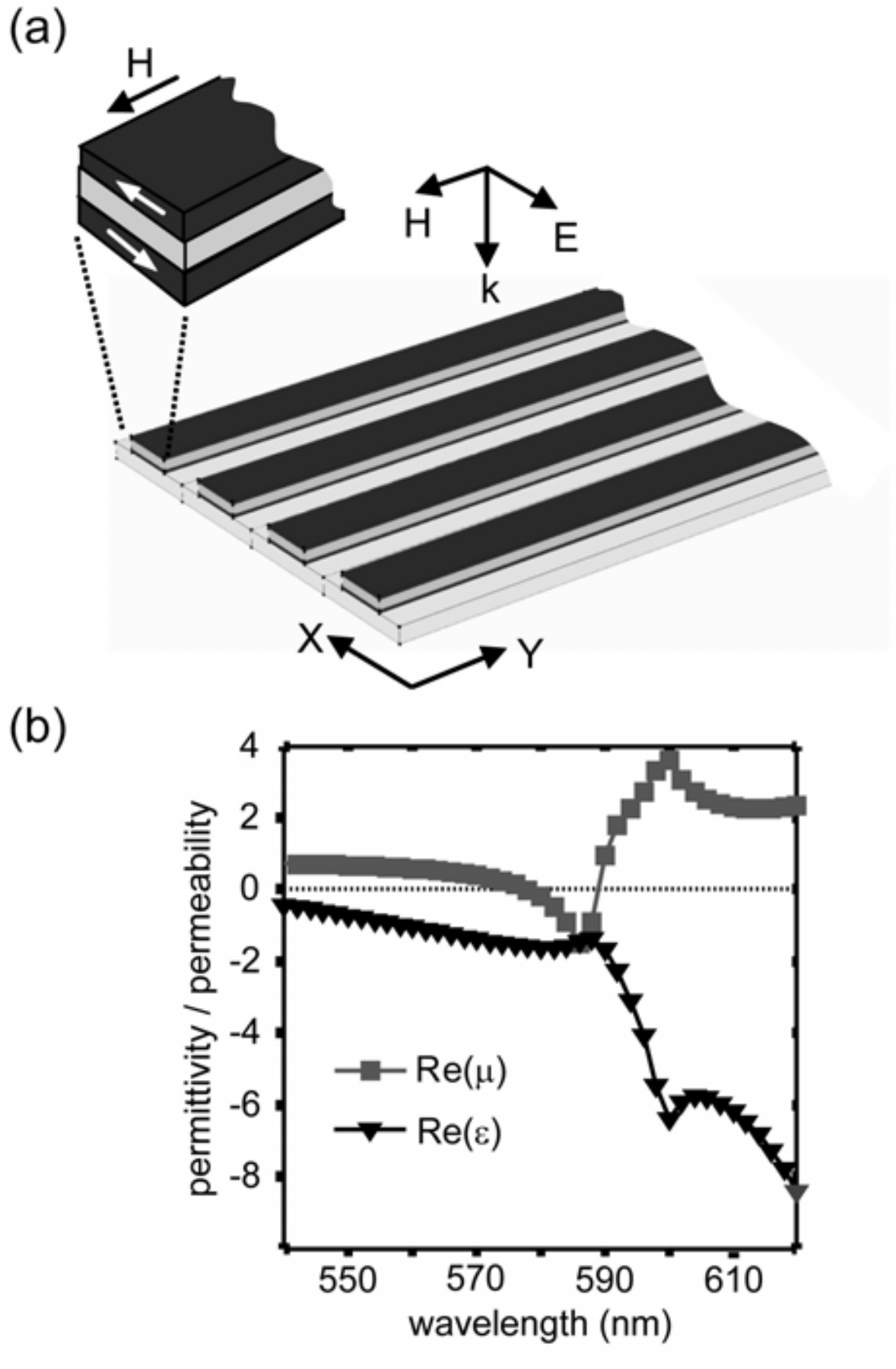

Fig. 8 (a) Double silver strips, separated by $Al_2O_3$. The strips are infinitely long in the y-direction and are periodically repeated in x-direction. The H field is oriented in the y-direction. Currents in both the strips are anti-parallel (white arrows in the magnified inset) if the H-field is polarized in y-direction. (b) Real parts of the permittivity and permeability as simulated with FEMFD.

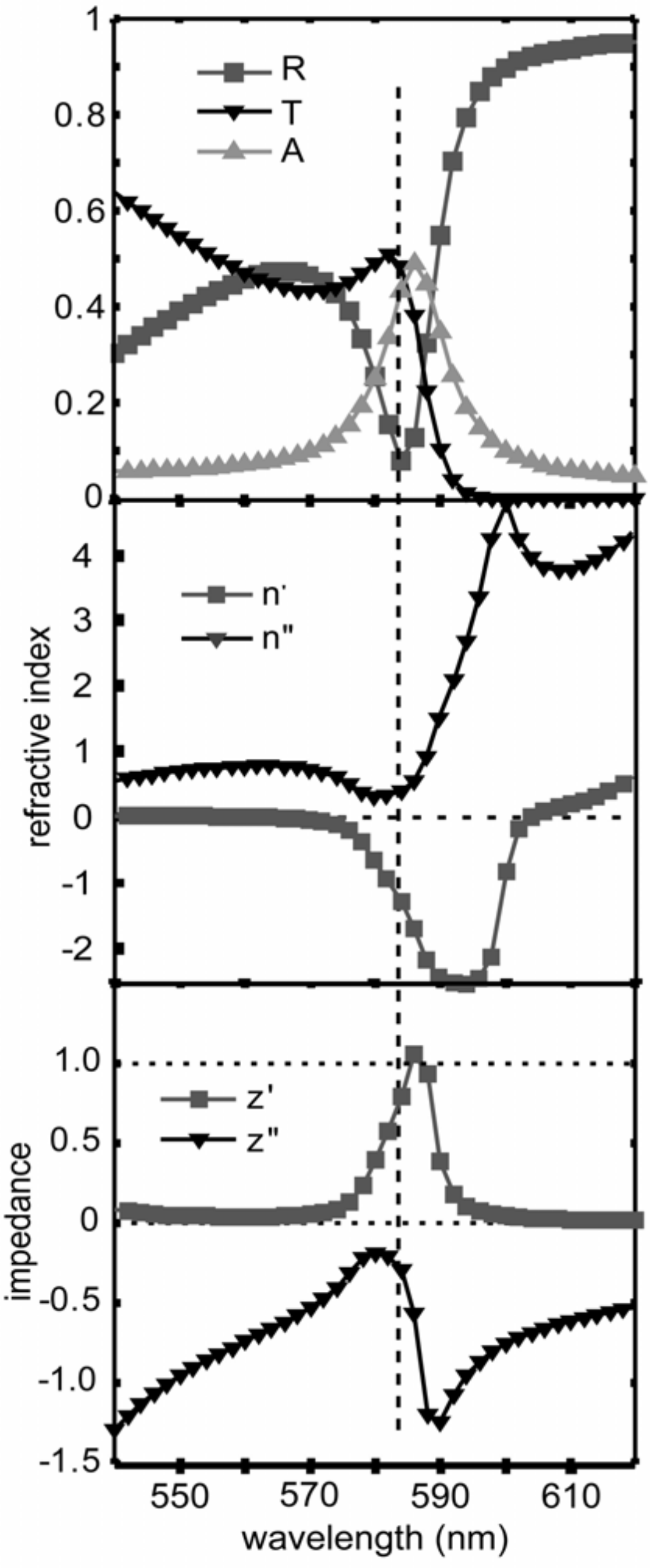

Fig. 9 Spectra of several optical constants of the structure shown in Fig. 8. Upper panel: Reflection R, transmission T, and absorption A spectra; middle panel: real and imaginary part of the refractive index; lower panel: real and imaginary part of the impedance. The vertical dashed line at 584 nm indicates a spectral region where the reflection is minimal, the transmission is high, the imaginary part of the refractive index is only 0.4 while the real part is negative and the real part of the impedance is close to 1, indicating impedance matching to air. The spectra were determined using FEMFD simulations.

calculations to determine the spectra of the electrodynamic constants. Fig. 8b shows the real parts of the permittivity (triangles) and of the permeability (squares). It is seen that both are negative at wavelengths between 580 and 590 nm.

The spectra of the reflectance, transmission, absorption, refractive index, and impedance are displayed in Fig. 9. It can be seen in Fig. 9a that the transmission has a local maximum of 51% at 582 nm. This is because the reflection has a local minimum and the absorption is limited. Indeed, the impedance is matched quite well from 582 to 589 nm, i.e. $Z' > 0.5$ and eventually reaching 1 at 586 nm, and simultaneously $|Z''| < 0.5$ in the range 570 - 585 nm (Fig. 9c). In total, this leads to a reflectance of less than 10% at 584 nm.

The absorption seems to have a local maximum at 586 nm, however it does not reproduce in the spectrum of $n''$. This is mainly because the reflection at the interface between air and the metamaterial hinders the electromagnetic radiation from entering the metamaterial at longer wavelengths and therefore the effective absorption of radiation inside the metamaterial is low for longer wavelengths. Still, it accounts for almost 90% of the losses in the range of the “reflectance window” at 584 nm. In summary of this section, we have shown that a metamaterial consisting of pairs of silver strips as depicted in Fig. 8a can form an almost impedance matched negative index material for visible light. The transmission is limited to 50% almost solely due to absorption, while reflection losses play a minor role.

## V. Gain, Compensating for Losses

It has recently been pointed out, that energy can be transferred from gain material to surface plasmon polaritons [44-49] or to plasmons in metal nanostructures [50, 51] using stimulated emission. Specifically, continuous thin films of metal were used to confine lasing modes in quantum cascade lasers to the gain region and also to guide the lasing modes by surface plasmon modes [44, 45]. Ramakrishna and Pendry suggested to staple gain materials such as semiconductor laser materials in between the negative index (or metal) layers of stacked near-field lenses [52] in order to remove absorption and improve resolution. The requirement of a perfect near-field lens, where thin layers of positive and negative index materials are alternated is, that $\varepsilon_P = -\varepsilon_N$ and simultaneously $\mu_P = -\mu_N$ where the subscripts denote materials constants of positive ($P$) and negative ($N$) materials. This requirement naturally includes the conditions $\varepsilon''_P = -\varepsilon''_N$ and $\mu''_P = -\mu''_N$, i.e. the positive layers must provide gain in order to optimize the lens [52].

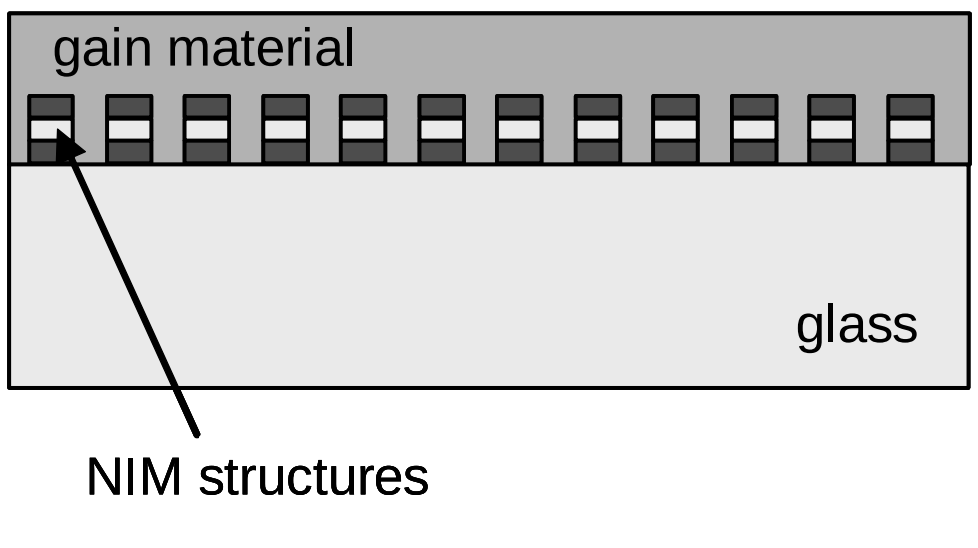

Fig. 10 Negative index material (e.g. double rods), filled with a gain medium, e.g. a solid solution of dye molecules in a matrix.

In our discussion we would like to turn to the refractive index rather than the permittivity and the permeability, because the absorption ($\alpha$) and gain ($g$) coefficients are more straightforwardly connected to the refractive index: $n'' = \frac{\lambda}{4\pi}(\alpha - g)$. Further, instead of alternating negative and positive index materials we propose to "submerge" the negative index structures (containing e.g. metal nanorods) in the gain media as shown in Fig. 10. For example, this could be achieved by spin coating a solution of laser dye molecules or π-conjugated molecules on top of the negative index structures. Applying semiconductor nanocrystals could be an alternative approach.

One might question whether the metal nanostructures nullify any attempt to amplify electromagnetic fields using gain materials in their close vicinity because gold nanoparticles are well known to quench fluorescence in an extremely efficient manner [53, 54]. In contrast, however, working solid state and organic semiconductor lasers show that sufficient gain can be provided so that in devices containing metal layers or metal nanoparticles the losses can be compensated. For instance, it was shown that an optically pumped organic laser comprising a metal-nanoparticle distributed feedback (DFB) grating needs only a marginally increased pumping threshold (compared to organic lasers with metal-free DFB gratings) to be operative [55]. In the case of infrared quantum cascade lasers (QCL), a wave guiding metallic layer was shown to be beneficial for the laser power output [45]. This astonishing result is due to an increased overlap of the surface plasmon guided mode profile with the gain region (the quantum cascade structure, in this case). This overlap offsets the increased losses (compared to a metal-free QCL) resulting from surface plasmon excitation. The net effect is an overall improved laser performance. We therefore conclude that it should indeed be feasible to use gain materials in order to compensate for the losses introduced by the resonant plasmonic metal nanoparticles in negative index metamaterials.

We want to give a specific example on the basis of the sample shown in Figs. 8 and 9. For the moment, we assume that the metal strips are submerged in a 200 nm thick layer of gain material (Fig. 10). We further assume that the gain material and the metal strips do not influence each other. This is an assumption that certainly needs to be discussed, but for the moment we shall assume that the gain of the material is not influenced by the metal strips. At the wavelength of least reflectance (due to impedance matching, $\lambda = 584\ \mathrm{nm}$), the strip material shows an absorption of approximately 45% (Fig. 9a). Applying Lambert-Beer's law and assuming that the absorptive loss should be fully compensated by the 400 nm thick gain layer, it turns out that a gain of $g = 3 \cdot 10^4\ \mathrm{cm}^{-1}$ is required. Let us further assume that we use Rhodamine 6G dissolved in some optically inert polymer. Rhodamine 6G has an stimulated emission cross section of $\sigma_{SE} = 3 \cdot 10^{-16}\ \mathrm{cm}^2$ [56] and therefore the concentration of excited dye molecules should be 170 mM. Alternatively, semiconductor nanocrystals (NC) such as CdSe NCs could be applied. It has been shown in [57] that the absorption cross section per NC volume can be as large as $10^5\ \mathrm{cm}^{-1}$. Because $g$ and $\alpha$ are usually of similar magnitude, we conclude, that densely packed nanocrystal films can show gain in the order of $g \approx 10^5\ \mathrm{cm}^{-1}$.

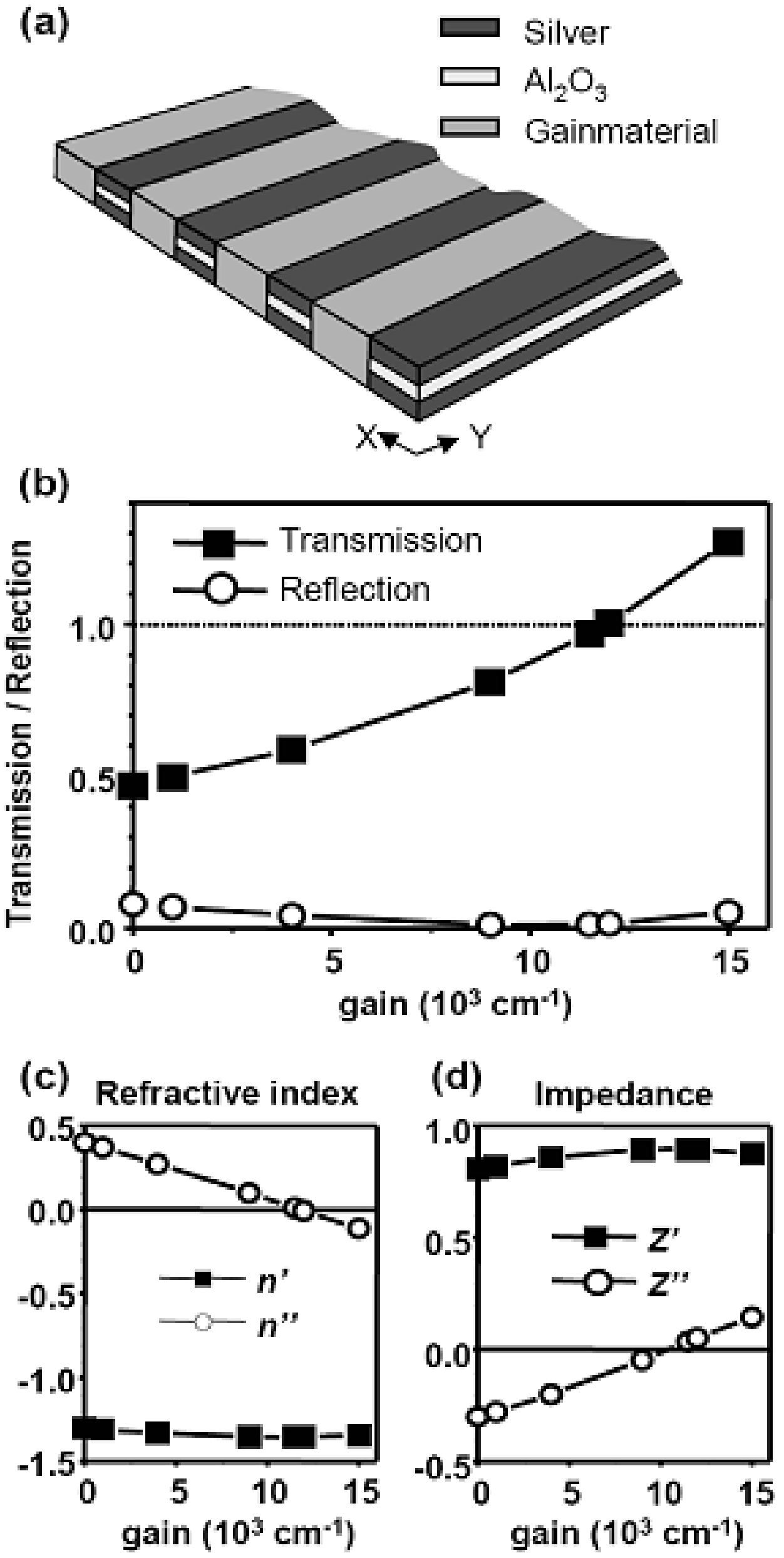

Fig. 11 (a) Same sample as in Fig. 8a, but with gain providing material in between the double silver strips. Air is assumed above and below the layer, and the layer is irradiated with a plain wave (584 nm) from above, H-field polarized along the y-direction. (b) Transmission and reflection as a function of the gain. At $g$ = 12 000 1/cm gain and losses cancel each other. Interestingly, the reflection shows also a minimum at $g$ = 12 000 1/cm. (c,d) refractive index and impedance as a function of gain. $n' \approx -1.35$ for all investigated gain levels. The spectra were determined using FEMFD simulations.

It is seen that the dye or nanocrystal concentrations need to be quite high to compensate for the losses. However, we have assumed in our rough estimation that the gain of the material in between the metal strips is not affected by the local fields in the vicinity of these metal strips. These fields can be quite high due to nano-plasmonic resonances. In fact, it has been pointed out by Kim et al. [58] and by Lawandy [51] that a gain medium and localized plasmonic resonances may lead to extremely high effective polarizabilities of the combined system. Therefore, the possibility may arise that each pair of gold nanorods as shown in Fig. 2, or each pair of strips as in Fig. 8 shows a much larger response to an incoming electric field as the same metal structure without gain.

In the example given above we have neglected that the gain material is in intimate contact with the silver strips. In order to get a better picture, we applied FEMFD simulations on the following model (Fig. 11a): We took the same structure as shown in figure 8, but now we filled the gaps in-between the double silver strips with a material that provides a fixed amount of gain between $0$ and $15\cdot10^{3}$ cm$^{-1}$. Fig. 11b-d show the transmittance ($T$), reflectance ($R$), refractive index ($n'$ and $n''$) and impedance ($Z'$ and $Z''$) as a function of gain ($g$). We found that with a gain of $12\cdot10^{3}$ cm$^{-1}$ the structure becomes transparent (Fig. 11b), while the real part of the refractive index $n'$ is almost unaffected by the gain material (Fig. 11c). Further more, the impedance which was already matched quite well without the gain medium (Fig. 9) improves further when gain is applied, i.e. $Z' \approx 1$ and $Z'' \approx 0$ for $g = 12\cdot10^{-3}\mathrm{cm}^{-1}$ (Fig. 11d). The exact results for a gain of $g = 12\cdot10^{-3}\mathrm{cm}^{-1}$ are $n' = -1.355$, $n'' = -0.008$, $Z' = 0.89$, $Z'' = 0.05$, $T = 100.5\%$, and $R = 1.6\%$.

Actually, if a critical magnitude of gain is surpassed, the polarizability and the field enhancement do not depend on nanoparticle shape or material any longer, but are solely limited by gain saturation in the gain medium [51]. At present, we have not included gain saturation in our model. It could be envisioned, that the gain material does not “simply” restore energy, which is lost due to absorption by the metal nanostructures, but it becomes an instrumental element of the negative index material, e.g. heavily increasing the negative response of the pairs of nanorods [51]. This will allow the design of negative index materials of less overall metal content. The density of pairs of rods may be reduced, or the size of each pair may be reduced, while the overall effective negative response of the metamaterial remains strong. This exciting field certainly needs more consideration, which will be given elsewhere.

## VI. Conclusion

Very recently, metamaterials have been designed that show a negative real part of the refractive index at the telecom wavelength of 1500 nm or 200 THz. Keeping in mind that it only took 5 years to come from 10 GHz up to 200 THz, we have no doubt that a negative refractive index metamaterial will be soon available also for the visible range. We have shown in numerical simulations that two key remedies are now available to overcome major obstacles that currently limit the development of optical negative-index materials (i) impedance matching designs are capable to suppress high reflectance, and (ii) gain materials embedded in metallic nanostructures can fully compensate for absorptive losses while still retaining the negative refractive index.

## Acknowledgment

We would like to acknowledge fruitful collaboration with V. A. Podolskiy, A. K. Sarychev, W. Cai, U. K. Chettiar, and H.-K. Yuan.